\documentclass[12pt]{article}

\usepackage{amsmath,amsfonts,epsfig,mathrsfs,todonotes,yfonts}
\usepackage{cancel}
\usepackage{xcolor}
\usepackage{cite}
\usepackage{stmaryrd}
\usepackage{mdframed}
\usepackage{scalerel}
\usepackage[top=2.5cm, bottom=2.5cm, left=2.75cm, right=2.75cm]{geometry} 
\usepackage{dsfont}
\usepackage[vcentermath]{youngtab}
\numberwithin{equation}{section}
\usepackage{hyperref}
 \hypersetup{
     colorlinks=true,
     linkcolor=black,
     filecolor=blue,
     citecolor=blue,      
     urlcolor=cyan,
     }
\usepackage[most]{tcolorbox}

\newcommand{\re}[1] {(\ref{#1})}

\newcommand{\pa}{\partial} 

\newcommand{\ber}{\begin{eqnarray}}
\newcommand{\eer}[1]{\label{#1}\end{eqnarray}}
\newcommand{\eero}{\end{eqnarray}}

\newcommand{\balg}{\begin{align}}
\newcommand{\ealg}{\end{align}}

\newcommand{\beq}{\begin{equation}}
\newcommand{\eeq}{\end{equation}}
\newcommand{\bea}{\begin{eqnarray}}
\newcommand{\eea}{\end{eqnarray}}

\newcommand{\rd}[1]{{\color{red}{#1}}}

\newcommand{\nn}{\nonumber}
\newcommand{\na}{\nabla}

\newcommand{\half}{{\textstyle{\frac12}}}
\newcommand{\ihalf}{{\textstyle{\frac i 2}}}

\def\HollowBox #1#2{{\dimen0=#1 \advance\dimen0 by -#2
       \dimen1=#1 \advance\dimen1 by #2
        \vrule height #1 depth #2 width #2
        \vrule height 0pt depth #2 width #1
        \llap{\vrule height #1 depth -\dimen0 width \dimen1} 
       \hskip -#2
       \vrule height #1 depth #2 width #2}}

\usepackage{fancyhdr} 
\newcommand{\auth}{\large Ulf Lindstr\"om ${}^{a,b}$\footnote{email: ulf.lindstrom@physics.uu.se}
and {\"O}zg{\"u}r Sar{\i}o\u{g}lu ${}^a$\footnote{email: sarioglu@metu.edu.tr}}
\begin{document}
\begin{flushright}
{\small UUITP-06/22}\\
\vskip 1.5 cm
\end{flushright}

\begin{center}
{\Large{\bf Tensionless Strings and Killing(-Yano) Tensors}}
\vspace{.75cm}

\auth
\end{center}
\vspace{.5cm}
\vspace{.5cm}
\centerline{${}^a${\it \small Department of Physics, Faculty of Arts and Sciences,}}
\centerline{{\it \small Middle East Technical University, 06800, Ankara, Turkey}}
\vspace{.5cm}
\centerline{${}^b${\it \small Department of Physics and Astronomy, Theoretical Physics, Uppsala University}}
\centerline{{\it \small SE-751 20 Uppsala, Sweden}}

\vspace{1cm}


\centerline{{\bf Abstract}}
We construct invariants for bosonic and spinning tensionless (null) strings in backgrounds that carry Killing tensors or Killing-Yano tensors of mixed type. This is facilitated by the close relation of these strings to point particles. We apply the construction to the Minkowski and to the Kerr-Newman backgrounds.

\bigskip

\noindent

\vskip .5cm
  
\vspace{0.5cm}
\small
\pagebreak
\tableofcontents

\renewcommand{\thefootnote}{\arabic{footnote}}
\setcounter{footnote}{0}

\section{Introduction}
In this note we point out that the construction of invariants based on the equation of motion for 
particles can be straightforwardly applied to tensionless strings\footnote{B. Carter once urged U.L. 
to use the nomenclature ``null strings" rather than tensionless strings, presumably thinking of, e.g., cosmic strings with higher-dimensional windings where there are solutions with zero effective tension in the Minkowski part \cite{Yamauchi:2014ita}. However, since for us they arise 
in relativistic strings in the limit that the tension goes to zero, ``tensionless'' is informative. In fact when supersymmetry 
is involved, it is not always obvious that the world sheet is null. Here we will use both descriptions 
interchangeably.}. This is because such strings can be viewed as 
collections of massless particles subject to certain constraints. 

Killing tensors, Killing-Yano forms and Killing-Yano tensors have a rich history of applications, 
e.g., to separation of variables in gravity, for finding symmetries of various differential operators 
\cite{Eastwood:2002su}, \cite{Howe:2016iqw}, to $G$-structures 
\cite{Papadopoulos:2007gf, Papadopoulos:2011cz}, and for finding 
geometric invariants\footnote{For recent discussions of conserved currents in this context see, e.g., 
\cite{Lindstrom:2021qrk}, \cite{Lindstrom:2021dpm}.}. Good general references are, e.g., 
\cite{Santillan:2011sh}, \cite{Hansen} and \cite{Chervonyi:2015ima}. For a very recent application, 
see \cite{Papadopoulos}. For the applications here, see \cite{Gibbons:1993ap} and 
\cite{Howe:2018lwu}.

Likewise there is by now a large literature on tensionless (null) strings. They were first introduced 
in \cite{Schild:1976vq} and independently rediscovered in a different guise in \cite{Karlhede:1986wb}. 
Relevant references here are \cite{Lindstrom:1990qb}, \cite{Lindstrom:1990ar} and 
\cite{Isberg:1993av}. In addition, the conformal string \cite{Gustafsson:1994kr} and a general 
discussion of various limits of branes and other gravitational systems \cite{Lindstrom:2000pp} 
may be consulted for a more general picture.

We first discuss the Killing tensor construction in the bosonic particle case in Sec.~\!\ref{BG}, then its 
extension to the tensionless bosonic string in Sec.~\!\ref{bosonic}. The tensionless spinning string 
and its relation to spinning particles is given in Sec.~\!\ref{spinning}. There the Killing tensors are 
generalized Killing-Yano ones with both symmetric and antisymmetric sets of indices.

\section{Background} \label{BG}
Given a spacetime ${\cal M}$ that admits a symmetric rank $n$ Killing tensor
\ber
\nabla_{(\mu}K_{\mu_1 \dots \mu_n)}=0~,
\eer{KT}
there is a quantity which is conserved along the geodesics of 
the geometry\footnote{See, e.g., \cite{Penrose:1986ca}.} $X^\mu(\tau)$:
\ber
\mathbb{K}=K_{\mu_1 \dots \mu_n}p^{\mu_1} \dots p^{\mu_n} \,,
\eer{21}
with
\ber
p^\mu=\pa_\tau X^\mu(\tau) := \dot{X}^{\mu}~.
\eer{}
So one has
\ber
\frac D {d\tau} p^\mu=0
\eer{Geo}
with $\frac D {d\tau}$ the directional derivative along the geodesic.
Conservation follows from 
\ber
\frac D {d\tau} \mathbb{K}=p^\mu\nabla_{\mu}\mathbb{K} = 
(p^\mu\nabla_{\mu}K_{\mu_1 \dots \mu_n})p^{\mu_1} \dots p^{\mu_n} + 
K_{\mu_1 \dots \mu_n}\frac D {d\tau} (p^{\mu_1} \dots p^{\mu_n})=0~,
\eer{}
where the first term vanishes due to the defining property of the Killing tensor \re{KT} and the second  
due to \re{Geo}. Note that this is a purely geometric construction assuming the existence of a 
Killing tensor and geodesics.

When we consider the relativistic massless particle action 
\ber
S_1=\half \int d\tau \dot X^\mu G_{\mu\nu}(X) \dot X^\nu~,
\eer{}
the equations of motion describe geodesics and the above construction yields a quantity $\mathbb{K}$ 
which is conserved on the motion.

If we try to extend the construction to the bosonic string, we run into difficulties since the equations 
of motion are not amenable to the same treatment. In \cite{Santillan:2011sh} the bosonic and the 
spinning strings are nevertheless treated in a similar way by imposing gauge conditions and 
constraints that effectively reduce them to particles.  Here we instead turn to their tensionless limits 
which are particle like by construction.

\section{The tensionless bosonic string}
\label{bosonic}

Tensionless strings (null strings) can be thought of as the high energy limit of ordinary strings. 
Here we investigate conserved currents for that limit.

Let $X$ be maps from the world sheet $\Sigma$ to the target space $\cal M$.
As described in, e.g., \cite{Isberg:1993av} the tensionless limit of the bosonic string may be 
described by the action
\ber
S_2={{\half}} \int d^2x V^aV^b\gamma_{ab}:=
{{\half}}\int d^2x~\! \pa{} X^\mu G_{\mu\nu}(X)\pa{}X^\nu~,
\eer{Act1}
where $V^a$ is a covariant world sheet density vector of weight one half, $\pa{} := V^a\pa_{a}$ 
and $\gamma$ is the induced metric
\ber
\gamma_{ab}=\pa_{a}X^\mu \pa_{b}X^\nu G_{\mu\nu}(X)~.
\eer{}
The field equation for $V^a$ is
\ber\label{null}
&&\delta V^a:~~~V^b\gamma_{ab}=0~,~~~\Rightarrow \det \gamma_{ab}=0~.
\eer{}
The variation of $X^\mu$ gives
\ber\nn
&&\delta X^\mu:~~~\int d^2x~\!\big(\pa (\delta X^\mu) G_{\mu\nu}\pa X^\nu+\half\pa X^\rho G_{\rho\nu},_\mu\pa X^\nu\delta X^\mu\big)=0\\[1mm]\nn
&&\Rightarrow \int d^2x~\!\tilde \na(\delta X^\mu G_{\mu\nu}\pa X^\nu )\\[1mm]
&&-\int d^2x~\!\delta X^\mu \big(G_{\mu\nu},_\rho\pa X^\rho\pa X^\nu+G_{\mu\nu}\tilde\na\pa X^\nu-\half\pa X^\rho G_{\rho\nu},_\mu\pa X^\nu \big)=0~,
\eer{invar}
where 
\ber
\tilde\nabla=V^a\tilde\nabla_{a}=\pa+\tilde \Gamma
\eer{}
 is a world sheet covariant derivative \cite{Isberg:1993av}. It satisfies  
 \ber
 \tilde\nabla_{a}V^a=0
 \eer{Vcond} 
 which determines
 the contracted world sheet connection to obey
 \ber
V^a\tilde \Gamma ^b_{~a b}=-2\pa_{a}V^a.
 \eer{metr}
 Using \re{metr} the first integral term in \re{invar} becomes a total derivative
 \ber
 \int d^2x~\!\pa_a(V^a\delta X^\mu G_{\mu\nu}\pa X^\nu )
 \eer{}
 and may be discarded. The remaining terms in \re{invar}  then imply
 \ber
 \tilde \na  \pa X^\mu+\pa X^\nu\Gamma^\mu_{~\nu\rho}\pa X^\rho=0~,
 \eer{FE}
where $\Gamma$ (without the tilde) denotes the Levi-Civita connection on $\cal M$.
 
The equations \re{null} and  \re{FE} mean that the world sheet spanned by the tensionless string 
is a null surface and that the string behaves classically as a collection of massless particles, one 
at each $\sigma$ position, constrained to move transversally to the direction of the string. This is clarified in 
appendix \ref{appc}.
\subsection{Invariances of the action}
\label{Invar}

We look for an invariance of the action modulo the equations \re{FE}. The field equations are 
derived assuming variations that vanish on the boundary. Suppose that $\delta X^\mu = K^\mu(X)$ 
leaves the action invariant but does not vanish on the boundary of the variation. Modulo the field 
equations, we are then left with
\ber\nn
\delta S_2&=&-\int d^2x~\! \tilde\na{} \Big(\delta X^\mu G_{\mu\nu}(X)\pa{}X^\nu\Big)\\[1mm]\nn
&=&
-\int d^2x~\! \Big(\tilde\na{}K_\nu \pa{}X^\nu
+K_\nu \tilde\na{}\pa{}X^\nu\Big)\\[1mm]\nn
&=&-\int d^2x~\! \Big(\pa{}_\mu K_\nu \pa X^\mu\pa{}X^\nu
-K_\rho \Gamma_{~\mu\nu}^{\rho}\pa{}X^\mu\pa{}X^\nu\Big)\\[1mm]
&=&-\int d^2x~\! \Big(\na{}_{(\mu}K_{\nu)}\pa X^\mu\pa{}X^\nu\Big)~,
\eer{Act3}
where we used the equation for $\tilde\na{}\pa X$ from \re{FE}. The expression for $\delta S_2$ is seen to 
vanish when $K$ is a Killing vector. There is a conserved quantity for each Killing vector 
of the background geometry. In fact, the weaker condition that $K$ is a conformal Killing vector
\ber
\na{}_{(\mu}K_{\nu)}=\lambda G_{\mu\nu}
\eer{}
is sufficient in view of  \re{null}.

It is interesting that the above argument can be extended to include higher rank Killing tensors. 
To show this, we first note that the action \re{Act1} is equivalent to the following phase space action
\ber
S_3=\int d^2x~\! \Big(-\half p_\mu p^\mu +p_\mu \pa X^\mu\Big)~.
\eer{Act4}
The field equations are now
\ber\nn
&&p^\mu=\pa X^\mu~,\\[1mm]\nn
&&p_\mu\pa_aX^\mu=0~,\\[1mm]
&& \na p^\mu =\tilde \na p^\mu +p^\nu\Gamma_{~\nu\rho}^{\mu}p^\rho=0~,
\eer{FE1}
where the first equation shows that $p_\mu$ is the ``momentum''
\ber
p_\mu=\frac {\pa {\cal L}}{\pa(\pa X^\mu)}
\eer{}
for the Lagrangian in \re{Act1}. Note that the equations \re{FE1} imply that $p^2=0$ on shell.

Introducing a phase space function 
\ber
{\cal K}^\mu(X,p)=K^\mu(X)+K^{\mu\nu}(X)p_\nu+K^{\mu\nu\rho}(X)p_\nu p_\rho+\dots~,
\eer{KEX}
we may repeat the derivation that led to \re{Act3}: Assume that $\delta X^\mu={\cal K}^\mu$ is 
a variation that does not vanish on the boundary but leaves \re{Act4} invariant. The variation 
of \re{Act4} leads to terms that vanish due to the equations \re{FE1} and leaves us with
\ber
\delta S_3=\int d^2x~\! \tilde\na \big( p_\mu {\cal K}^\mu\big)~.
\eer{Var}
Using the field equations \re{FE1}, it is easy to see that this vanishes term by term when the 
expansion coefficients in \re{KEX} are Killing tensors \re{KT} or conformal Killing tensors: 
\ber
\nabla_{(\mu}K_{\mu_1 \dots \mu_n)}=
G_{(\mu\mu_1}\na_{|\lambda |} K^\lambda_{~\mu_2\dots \mu_n)}~.
\eer{CKT}

At this point we may invoke a version of Noether's theorem to conclude that the following is 
annihilated by $\tilde\na$ on-shell 
\ber
J= \frac {\pa {\cal L}}{\pa(\pa X^\mu)}{\cal K}^\mu~=G_{\mu\nu}\pa X^\nu {\cal K}^\mu=p_{\mu} {\cal K}^\mu~~~
\Rightarrow ~~~~\tilde \na J=0~.
\eer{tauinv}
Unlike the usual Noether procedure, where time is the only variable, this does not mean that 
$J$  is constant since it is only shown to be independent of one combination of the world sheet 
coordinates. To proceed, we write
\ber
0=\tilde \na J=V^a\tilde\na_a J=\tilde\na_a (V^a J):=\tilde\na_a J^a 
= \tilde\na_\tau J^\tau+\tilde\na_\sigma J^\sigma~,
\eer{}
which states that $J^a$ is (world sheet) divergence free. Here we have used 
$\tilde\nabla_{a}V^a=0$ and denoted the world sheet coordinates by $(\tau,\sigma)$. 
Integrating over $\sigma$ and using the divergence theorem, we have that
\ber
\frac d {d \tau} \int d\sigma J^\tau=\int d\sigma \tilde\na_\tau J^\tau=-\int d\sigma  \tilde\na_\sigma J^\sigma=0~,
\eer{JJJ}
where the last equality is also obvious  in a world sheet diffeomorphism gauge\footnote{The available gauges for tensionless strings in this formulation are discussed in \cite{Isberg:1993av}.} where 
$(V^\tau,V^\sigma)=(v,0)$ with $v$ a constant. 

The interpretation as particles suggests that we can construct invariants directly along the same 
lines as for particles using Killing tensors and momenta. Since the derivation runs parallel to the
discussion about invariances of the Lagrangian, we relegate this to Appendix \ref{PL}. 

So we have shown that a bosonic tensionless string, subject to the relations \re{null}, \re{FE}, in a 
geometry that allows Killing tensors, has invariants that are $\sigma$ integrals of the expression 
in the expansion of $p_\mu {\cal K}^\mu$ using \re{KEX}.

Minkowski space has the maximal number of Killing vectors possible. The 15 (conformal) Killing 
vectors of the $D=4$ Minkowski space are:
\ber
\makebox{Translations $P_\mu$:}&&\eta^{\mu\nu}{\bf e}_\nu~,\\[1mm]
\makebox{Rotations in the ${\mu\nu}$-plane $L_{\mu\nu}$:}&&X_{[\mu}{\bf e}_{\nu]}~,\\[1mm]
\makebox{Dilatations $S$:}&&X^\mu{\bf e}_\mu~,\\[1mm]
\makebox{Special Conformal $K_\mu$:}&&2X_\mu X^\nu{\bf e}_{\nu}-X\cdot X{\bf e}_{\mu}~.
\eer{list}
The corresponding generators are listed in the left column and ${\bf e}_{\mu}$ denote the 
coordinate basis vectors.

The Killing vectors $V^{\mu}$ from \re{list} give the following invariants $\cal{Q}$ \re{JJJ}:
\ber\label{moment}
&&{\cal{Q}}^P_\nu=\int d\sigma p_\mu\delta^\mu_\nu=\int d\sigma p_\nu~,\\[1mm]\label{anmom}
&&{\cal{Q}}^L_{\mu\nu}=\int d\sigma p_{[\nu}X_{\mu]}~,\\[1mm]
&&{\cal{Q}}^S=\int d\sigma p_{\mu}X^{\mu}~,\\[1mm]
&&{\cal{Q}}^K_\mu=\int d\sigma\big(2(p\cdot X)X_\mu-X\cdot X p_\mu\big)~.
\eer{Minvar}
In the gauge $V^a=(v,0)$, the first equation in \re{null} becomes \re{cstr}
\ber
\dot X^2=\dot X X'=0~,
\eer{cstr1}
and, on shell, \re{moment} becomes
\ber
{\cal{Q}}^P_\nu=P_\nu:=\int d\sigma \hat p_\nu= v\int d\sigma \dot X_\nu~,
\eer{pi}
where 
\ber\hat p_\nu=\frac {\pa{\cal L}}{\pa\dot X^\nu}
\eer{hpi}
is the usual momentum. The relations \re{cstr1}--\re{hpi} are precisely the coordinate choices 
made in \cite{Schild:1976vq} where the total momentum \re{pi} is shown to be invariant. 
Similarly, \re{anmom} represents angular momentum in this gauge. All the relations 
\re{moment}--\re{Minvar} are direct extensions of those for massless particles.

All the higher Minkowski space Killing tensors are reducible, i.e., sums of products of Killing 
vectors, except the metric. However the integrand involving the metric vanishes on shell 
(for all geometries). We therefore turn to the more interesting example provided by the 
Kerr-Newman metric in $D=4$: The metric and the vector potential are given by
\bea
d s^{2} & = & - \left( \frac{\Delta - a^{2} \sin^{2} \theta}{\Sigma} \right) d t^{2} 
  - \frac{2 a \sin^{2}\theta \left( r^{2} + a^{2} - \Delta \right)}{\Sigma} \, d t \, d \phi \nn \\
& & + \left( \frac{\left( r^{2} + a^{2} \right)^{2} - \Delta \, a^{2} \sin^{2} \theta}{\Sigma} \right) 
\sin^{2}\theta \, d \phi^{2} + \frac{\Sigma}{\Delta} \, d r^{2} + \Sigma \, d\theta^{2} \,,
\label{KerrNewman} \\
A_{a} \, dx^{a} & = & - \frac{q r}{\Sigma} \left( dt - a \sin^{2}\theta \, d\phi \right) \,,
\eea
where
\beq
\Sigma = r^{2} + a^{2} \cos^{2}\theta \qquad \mbox{and} \qquad
\Delta = r^{2} + a^{2} +q^{2} -2 M r \,. \label{SigDel}
\eeq
The nontrivial Killing tensor for the Kerr-Newman metric has components
\ber
K_{tt} & = & \frac{\left(a^3 \cos{2 \theta}+a^3+2 a r^2\right)^2 \left(r^2 \sin^2{\theta}
+ \Delta \cos^2{\theta} 
 \right)}{4 \Sigma^3} ~,\nn \\
K_{t\phi} & = & -\frac{a \sin^2{\theta} \left(a^2 \cos{2 \theta}+a^2+2 r^2\right)^2 
\left(a^2 r^2+a^2 \Delta \cos^2{\theta} +r^4\right)}{4 \Sigma^3} ~,\nn \\
K_{rr} & = & -\frac{a^2 \cos^2{\theta} \left(a^2 \cos{2 \theta} + a^2+2 r^2\right)^2}{4 \Delta
 \Sigma } ~, \\
K_{\theta\theta} & = & \frac{r^2 \left(a^2 \cos{2 \theta}+a^2+2 r^2\right)^2}{4 \Sigma } ~,\nn \\
K_{\phi\phi} & = & \frac{\sin^2{\theta} \left(a^2 \cos{2 \theta}+a^2+2 r^2\right)^2 \left(a^4 \Delta  
\sin^2{\theta} \cos^2{\theta} +r^2 \left(a^2+r^2\right)^2\right)}{4 \Sigma^3} ~. \nn
\eer{KNKT}
One has $p^{\mu} = \partial (t,r,\theta,\phi)^{\mu}$ and thus a $\tau$ invariant \re{Inv} is given 
by $c(\sigma)=K_{\mu\nu} p^{\mu} p^{\nu}$, which reduces to the classical ``Carter constant" \cite{Carter} 
\emph{at each $\sigma$} when $V^a = (v,0)$. We thus find an invariant which is the integral of these
\ber
C=\int d\sigma c(\sigma)~.
\eer{}

\section{The tensionless spinning string}
\label{spinning}
In this section we extend the construction of Sec.~\!\ref{bosonic} to spinning strings in a flat 
background. 

The tensionless limit of the spinning string with an $O(N)$ symmetry introduced in 
\cite{Lindstrom:1990ar} has the Lagrangian 
\ber
2{\cal L} = (V^a\pa_{a}X^\mu+i\lambda^\mu_i\chi_{i})(V^b\pa_{b}X_\mu+i\lambda_{i\mu}\chi_i)
+ i\lambda^\mu_i V^a\pa_{a} \lambda_{i\mu} + A_{ij}\lambda^{\mu}_i\lambda_{j\mu}~.
\eer{}
Here $\lambda$ and $\chi$ are spinorial\footnote{Spinors are just Grassmann numbers in this formulation.} 
world sheet densities and $A$ is antisymmetric in the $O(N)$ indices $i,j$. Gauge 
fixing the world sheet diffeomorphisms displays the model as a set of spinning HPPT particles 
\cite{Brink:1976uf}, \cite{Howe:1988ft}. For our purpose we take $N=1$ and thus consider
\ber
2{\cal L}_C = (\pa X^\mu+i\lambda^\mu\chi)(\pa X_\mu+i\lambda_{\mu}\chi)
+ i\lambda^\mu \pa \lambda_{\mu} ~,
\eer{Lag}
where we used the definition $\pa=V^a\pa_a$. One feature of the complete set of field equations is that, 
on shell,
\ber
(\pa X^\mu+i\lambda^\mu\chi)(\pa X_\mu+i\lambda_{\mu}\chi)=0~.
\eer{onshell}
We will need a phase space formulation and define
\ber
p_\mu=\frac {\pa {\cal L}_C} {\pa( \pa X^\mu)}=\pa X_\mu+i\lambda_{\mu}\chi~.
\eer{}
Using this we find the equivalent $(X,p,\lambda)$ phase space Lagrangian\footnote{The 
covariant derivative $\tilde \na$ on the density $\lambda$  reduces to $\pa$ in the Lagrangian 
due to the Grassmann property of $\lambda$.}
\ber
{\cal L}_P=-\half p^2+p^\mu(\pa X_\mu+i\lambda_{\mu}\chi) 
+ \ihalf \lambda^\mu\tilde \na\lambda_\mu \,,
\eer{first}
where we suppressed the Minkowski metric in the contractions. The corresponding action is invariant 
under local supersymmetry transformations given in Appendix \ref{appb}.
The equations of motion are
\ber\nn
&\delta X^\mu&:~\tilde \na p_\mu=0\\[1mm]\nn
&\delta p^\mu&:~ p_\mu=\pa X_\mu+i\lambda_{\mu}\chi\\[1mm]\nn
&\delta \lambda^\mu~&:~ p_\mu\chi+\tilde \na\lambda_\mu=0\\[1mm]\nn
&\delta V^a~&:~ p_\mu\pa_aX^\mu+\tfrac{i}{2}\lambda_\mu\pa_a\lambda^\mu=0\\[1mm]
&\delta \chi~&:~ p_\mu\lambda^\mu=0~.
\eer{eom}
On shell we have \re{onshell}
\ber
p_\mu p^\mu=0~.
\eer{nulll}


In the partial gauge choice $(V^0,V^1)=(e^{-1/2},0)$, the Lagrangian \re{first} becomes
\ber
{\cal L} =-\frac e 2 \hat p^2+\hat p^\mu(\dot X_\mu+i\hat \lambda_{\mu}\hat\chi) 
+ \ihalf \hat\lambda^\mu \dot{\hat\lambda}_\mu~.
\eer{first2}
after the redefinitions
\ber
p^\mu=e^{1/2}\hat p^\mu~,~~~
\lambda^\mu=e^{1/4}\hat \lambda^{\mu}
~,~~~\chi=e^{-3/4}\hat \chi~.
\eer{}
 

As a result of this exercise, we see that an alternative formulation 
of the spinning null string is given by the Lagrangian \re{first2}, but with all fields depending on 
the additional coordinate $\sigma$ and a constraint that corresponds to integrating out $V^a$ 
in \re{first}. The latter reads
\ber
p_\mu\pa_aX^\mu+\tfrac{i}{2}\lambda_\mu\pa_a\lambda^\mu=e^{1/2}\Big(\hat p_\mu\pa_aX^\mu+\tfrac{i}{2}\hat\lambda_\mu\pa_a\hat \lambda^\mu\Big)=0~.
\eer{orth}

We have thus recovered in first order form the fact  that the tensionless spinning string in a 
particular gauge is equivalent to a bunch of spinning particles, one at each $\sigma$, moving 
under the orthogonality constraint \re{orth} (see Appendix \ref{appc}). We now note that the Lagrangian \re{first2} is formally equivalent to that of a spinning particle \cite{Howe:1988ft} and  we want to use this analogy to construct invariants 
for the tensionless spinning  string based on the Lagrangian \re{first2}, mimicking the construction for the spinning particle \cite{Brink:1976uf}, \cite{Howe:2018lwu}.

\subsection{Invariants for the spinning particle}
In this subsection we review the analysis of \cite{Howe:2018lwu}. The Lagrangian \re{first2} is taken 
to represent the spinning particle, so now there are only $\tau$ dependent fields. There is a natural 
symplectic form $\omega$ associated with this Lagrangian:
\ber
\omega=dX^\mu\wedge dp_\mu - \ihalf d\lambda^\mu\wedge d\lambda_\mu\ .
\eer{}
The corresponding Poisson bracket is
\ber
\{F,G\}=\frac{\pa F}{\pa X^\mu}\frac{\pa G}{\pa p_\mu}-\frac{\pa F}{\pa p_\mu}\frac{\pa G}{\pa X^\mu}
+i(-1)^f \frac{\pa F}{\pa\lambda^\mu}\frac{\pa G}{\pa\lambda_\mu}=(-1)^{fg+1}\{G,F\}\,.
\eer{}
With
\ber
H= \half p^2~,\quad Q= \lambda\cdot p \,,
\eer{align}
we have
\ber
\{X^\mu,p_\nu\}=\delta^\mu_\nu\ ,\qquad  \{\lambda^\mu,\lambda^\nu\}=-i\eta^{\mu \nu}\, 
\quad \Rightarrow ~~\{Q,Q\}&=-2iH~.
\eer{}
We may express the local supersymmetry of \re{first} using the Poisson brackets: 
\ber\nn
&& \{Q,X^\mu\}=-\lambda^\mu\\[1mm]\nn
&&\{Q,\lambda^\mu\}=-ip^\mu\\[1mm]
&& \{Q,p^\mu\}=0 ~.
\eer{}
In addition, there are the local supergravity transformations
\ber\nn
&&\delta e =- \epsilon \chi\\[1mm]
&&\delta\chi = i\dot\epsilon~.
\eer{}
The phase space invariants based on this Lagrangian involve (generalized) superconformal 
Killing-Yano tensors, as discussed in \cite{Howe:2018lwu}. Here we recapitulate this in the 
present setting.

In \cite{Howe:2018lwu}, a generalization of conformal Killing-Yano (CKY) forms to mixed 
symmetry conformal Killing-Yano tensor (CKYT) $A_{p,q}$ of type $(p,q)$ was introduced. 
This is a traceless mixed tensor with the Young tableau
\ber
A_{p,q}\sim\ \ \ \ \ {\overbrace{\young(\hfil\hfil\hfil\hfil\hfil,\hfil,\hfil,\hfil)}^{q}}
\eer{4.10}
with $(p+1)$ boxes in the first column. The differential constraint satisfied by such a CKYT is that, 
when a derivative is applied to $A_{p,q}$, the traceless tensor corresponding to the Young tableau 
with one extra box on the first row has to vanish, i.e.,
\ber
\pa A_{p,q}\ni \overbrace{\young(\hfil\hfil\hfil\hfil\hfil,\hfil,\hfil,\hfil)}^{q+1}\ = 0~.
\eer{4.11}

Such tensors appear naturally in the context of the spinning particle in \cite{Gibbons:1993ap},
albeit there in the context of global supersymmetry in curved backgrounds. 

A function on phase space, of the form\footnote{C.f. \re{21}.}
\ber
{\cal F}=F(X,\lambda)^{\mu_1\ldots \mu_q}p_{\mu_1}\ldots p_{\mu_q}
\eer{}
can be expanded in the odd variables to give a sum of terms of the following kind
\ber
\lambda^pA_{p,q}p^q:=\lambda^{\nu_1\ldots \nu_p} A_{\nu_1\ldots \nu_p,\mu_1\ldots \mu_q}\, 
p^{\mu_1 \ldots \mu_q}\ ,
\eer{}
where the multi-index $\lambda$ and $p$ denote $p$-fold and $q$-fold products of the odd 
coordinates and the even momenta, respectively. A world line super-invariant is a function $F$
of the phase-space variables which is weakly annihilated by $Q$, $\{Q,{\cal F}\}\approx 0$. Since 
$\{Q,Q\}\sim H\sim \frac{d}{d \tau}$ such a function will automatically be a constant of the motion 
modulo in the particle case. If we are given such an invariant function with lowest $\lambda$ component 
$\lambda^pA_{p,q}p^q$, then we may use the invariance condition $\{Q,{\cal F}\}\approx 0$ to determine 
the rest of the $\lambda$ components. The result is
\ber
{\cal F}=\lambda^p A_{p,q}p^q+\alpha(p,q)\lambda^{p+2} dA_{p+2,q-1} p^{q-1}:=A+dA \,,
\eer{Finv}
where $A_{p,q}$ is in the representation \re{4.10}, satisfies the constraint \re{4.11} and
\ber
\alpha(p,q):= i\frac{(-1)^{(p+1)} q}{(1+p+q)}\ ,
\eer{}
\ber
\left(dA_{p+2,q-1}\right)_{\nu_1\ldots \nu_{p+2},\mu_1\ldots \mu_{q-1}}
:=\pa_{[\nu_1}A_{\nu_2\ldots \nu_{p+1},\nu_{p+2}]\mu_1\ldots \mu_{q-1}}~.\ 
\eer{}

\subsection{Application to the tensionless spinning string}
To use these results for the tensionless spinning  string, we replace the time derivative in the above 
construction by $\pa$, make a final gauge fixing $V^a\to (1,0)$ and observe that ${\cal F}$ in \re{Finv} obeys 
\ber
\tilde\na {\cal F}=0
\eer{}
on shell, i.e. using the field equations that follow from \re{first2}. These are
\ber\nn
&p^2=0~&\\[1mm]\nn
&\tilde \na p_\mu=0~&\to \dot p_\mu=0\\[1mm]\nn
&\tilde\na \lambda_\mu+i\chi\lambda_\mu=0~&\to \dot \lambda_\mu=0\\[1mm]\nn
&p_\mu=e^{-1}\big(\pa X_\mu-i\chi\lambda_\mu\big)~&\to p_\mu=\dot X_\mu\\[1mm]
&\lambda p=0~,&
\eer{Feq2}
where the right hand sides are the fully gauge fixed relations, including $\chi = 0$.
The same arguments as in the bosonic case in Sec.~\!\ref{bosonic} now ensure that 
\ber
\frac d {d\tau}\int {\cal F} d\sigma =0
\eer{}
on the motion.

A simple example of these invariants is given by the $\tau$-invariant with leading term $A_{0,1}$:
\ber
{\cal F}=A,_\mu p^\mu-\ihalf \lambda^{\nu_1}\lambda^{\nu_2}\pa_{{\nu}_1}A,_{{\nu}_2}~.
\eer{}
It is not difficult to see that $\tilde\na {\cal F}=\dot {\cal F}=0$ using \re{Feq2} and $\pa_{(\mu}A,_{\nu)}=0$.

When the leading term is instead $A_{1,1}$ we have
\ber\nn
&&{\cal F}=\lambda^\nu A_{\nu,\mu} p^\mu+{\textstyle \frac i 3}\lambda^{\nu_1}\lambda^{\nu_2}\lambda^{\nu_3}\pa_{[\nu_1}A_{\nu_2,\nu_3]}~.\\[1mm]\nn
&&\Rightarrow \tilde \na F=i\lambda^\nu\chi A_{\nu,\mu} p^\mu+\lambda^\nu\pa_\rho A_{\nu,\mu} (e p^\rho +i\chi\lambda^\rho)p^\mu+\chi\lambda^{\nu_1}\lambda^{\nu_2}\lambda^{\nu_3}\pa_{[\nu_1}A_{\nu_2,\nu_3]}\\[1mm]\nn
&&+\frac i 3\lambda^{\nu_1}\lambda^{\nu_2}\lambda^{\nu_3}\pa_\rho \pa_{[\nu_1}A_{\nu_2,\nu_3]}(e p^\rho +i\chi\lambda^\rho)~\\[1mm]
&&\to \lambda^\nu\pa_\rho A_{\nu,\mu}  p^\rho p^\mu +\frac i 3\lambda^{\nu_1}\lambda^{\nu_2}\lambda^{\nu_3}\pa_\rho \pa_{[\nu_1}A_{\nu_2,\nu_3]} p^\rho =0~,
\eer{}
where we used $\pa_{(\rho }A_{|\nu,|\mu)}=0$ and $\pa_{\nu_1}\pa_{(\rho}A_{|\nu_2,|\nu_3)}=0$ to 
show that the last line vanishes.

These arguments extend to arbitrary ${\cal F}$.

\section{Conclusions}
In this note we point out how to construct new invariants for tensionless strings, covering 
the tensionless bosonic  (null) string in a curved background and the tensionless spinning (null) string in flat background. 
Obvious further work would be to consider the spinning tensionless (null) string in curved background and to 
discuss the zero tension superstring. The latter has the flat super space action \cite{Lindstrom:1990qb}
\ber
S_{\rm{super}}=\int d^2xV^aV^b\Pi^\mu_a\Pi_{b\mu}
\eer{}
which is \re{Act1} with the space time supersymmetrisation
\ber
\pa_aX^\mu\to\Pi^\mu_a=\pa_aX^\mu-i\bar\theta\Gamma^\mu\pa_a\theta~,
\eer{}
where $\theta$ is the antisymmetric spinor partner to $X^\mu$. This system should be open to 
the same analysis as the tensionless bosonic (null) string. However, there is the added complication in curved 
supergravity backgrounds of defining super Killing-Yano tensors.

The tensionless strings may be viewed as high energy limits of ordinary strings 
\cite{Isberg:1993av} and hence the invariants discussed in this note should be relevant for 
ordinary strings in that limit \cite{Gross:1988ue}.

In the zero tension limit the strings lose their world sheet Weyl invariance but gain ambient 
space time conformal invariance. In the present note, this is partly reflected in allowing for 
conformal Killing vectors, but in looking for the most general conserved quantities that should 
be more systematically taken into account. A good setting for that would be the conformal string described in
\cite{Gustafsson:1994kr}.\\

\noindent{\bf Acknowledgments}\\
We are grateful to George Papadopoulos for comments on the manuscript.
The research of U.L. is supported in part by the 2236 Co-Funded 
Scheme2 (CoCirculation2) of T\"UB{\.I}TAK (Project No:120C067)\footnote{\tiny However 
the entire responsibility for the publication is ours. The financial support received from 
T\"UB{\.I}TAK does not mean that the content of the publication is approved in a scientific 
sense by T\"UB{\.I}TAK.}. 
\bigskip

\appendix
\section{Particle-like  invariants}
\label{PL}
Consider the quantity
\ber
\mathbb{K}=K_{\mu_1 \dots \mu_n}p^{\mu_1} \dots p^{\mu_n}~,
\eer{Inv}
where $\mathbb{K}$ is a function of $X$ (not a density) and $K$ is a Killing tensor of rank $n$
satisfying \re{KT} or a conformal Killing tensor \re{CKT}. Then the following world sheet 
covariant derivative of $\mathbb{K}$ vanishes:
\ber\nn
\tilde \na{}\mathbb{K} &= & \pa{}K_{\mu_1 \dots \mu_n}p^{\mu_1} \dots p^{\mu_n} + 
n K_{\mu_1 \dots \mu_n} (\tilde \na{}p^{\mu_1}) \dots p^{\mu_n}\\[1mm]\nn
& = &\nabla_{\mu}K_{\mu_1 \dots \mu_n}\pa{}X^\mu p^{\mu_1} \dots p^{\mu_n} + 
n \Gamma_{~\mu\mu_1}^{\lambda}K_{\mu_2 \dots \mu_n\lambda}\pa{}X^\mu p^{\mu_1} \dots p^{\mu_n}\\[1mm]
&& - ~\!n K_{\mu_1 \dots \mu_n}
(p^\sigma \Gamma_{~\sigma\rho}^{\mu_1}p^\rho)p^{\mu_2} \dots p^{\mu_n} = 0~,
\eer{}
where we have used the field equation \re{FE1} and the definition of covariant derivative 
in getting to the second line.  The first term vanishes due to \re{FE1} and \re{KT} or \re{CKT}, 
the remaining two cancel after using \re{FE1} once more. 

Note that, as before, this does not mean that $\mathbb{K}$ is constant since it is only shown 
to be independent of one combination of the world sheet coordinates. If we again go to a world 
sheet diffeomorphism gauge where $V^a=(V^\tau,0)$, the $\tau$ derivative of $\mathbb{K}$ 
vanishes but not the $\sigma$ derivative. However, if we again use\footnote{This is the condition  
allowing partial integration found in \cite{Isberg:1993av}.} $\tilde\nabla_{a}V^a=0$, we may write 
\ber
0=\tilde \na{}\mathbb{K} =V^a\tilde \na{}_a\mathbb{K}=\tilde \na{}_a(V^a\mathbb{K})~,
\eer{}
so that the construction in Sec. \ref{Invar} applies
\ber
\pa_\tau(V^\tau\mathbb{K})+\pa_\sigma(V^\sigma\mathbb{K})=0~.
\eer{}
Treating this as the conservation of the current $V^a\mathbb{K}$, we integrate over $\sigma$:
\ber
\int \pa_\tau(V^\tau\mathbb{K})d\sigma=\frac d {d\tau}\int (V^\tau\mathbb{K})d\sigma=
-\int \pa_\sigma(V^\sigma\mathbb{K})d\sigma=0~.
\eer{}
The last equality follows after specifying the boundary of the $\sigma$ integral, or by choosing the 
gauge $V^a=(V^\tau,0)$.

\section{SUSY transformations}
\label{appb}
Here are the local supersymmetry transformations that leave the action for \re{first} invariant:
\ber\nn
&&\delta X^\mu=i\epsilon\lambda^\mu\\[1mm]\nn
&&\delta p_\mu=i\epsilon\tilde \na \lambda_\mu\\[1mm]\nn
&&\delta \lambda^\mu=-\epsilon(\pa X^\mu+\ihalf \lambda^\mu\chi)\\[1mm]\nn
&&\delta V^a=iV^a(\epsilon\chi)\\[1mm]
&&\delta \chi=\tilde \na\epsilon~.
\eer{}
\section{The tensionless string as a collection of massless particles}
\label{appc}

As an alternative to the covariant treatment presented above, we may from the outset 
choose a world sheet diffeomorphism gauge
\ber
V^a = (v,0)~,
\eer{}
with $v$ a constant, corresponding to the conformal gauge in the tensile theory. We are then 
looking at the action
\ber
\int d^2x~v^2 \, \dot X^\mu G_{\mu\nu}(X)\dot X^\nu~,
\eer{GAct}
which leads to the geodesic equation
\ber
\ddot X^\mu+\Gamma_{~\rho\nu}^{\mu}\dot X^\rho\dot X^\nu=0~,
\eer{}
subject to the constraints from $V^a$ variation\footnote{C.f. the vanishing of the energy 
momentum tensor for the tensile string.} \re{null}
\ber
\dot X^2=\dot X X'=0~.
\eer{cstr}
This brings out the physical interpretation of the bosonic null string as a collection of massless 
particles moving along null geodesics orthogonal to the $\sigma$ direction of the string.

\end{document}